\documentclass[aps,preprint]{revtex4-1}
\usepackage{graphicx}
\usepackage{epstopdf}
\usepackage{bm}
\usepackage{lineno}

\usepackage[utf8]{inputenc}
\usepackage{amsmath,amssymb,lmodern} 

\newcommand{\be}{\begin{equation}}
\newcommand{\ee}{\end{equation}}
\newcommand{\nn}{\mbox{} \nonumber \\ \mbox{} }
\newcommand{\ba}{\begin{eqnarray}}
\newcommand{\ea}{\end{eqnarray}}
\newcommand{\om}{\omega}
\newcommand{\Alfven}{Alfv\'{e}n }

\newcommand{\E}{{\bf E}}
\newcommand{\B}{{\bf B}}

\newcommand{\Bf}{{magnetic field}}
\newcommand{\Ef}{{electric field}}

\newcommand{\mss}{magnetospheres}

\newcommand{\Sch}{Schwarzschild}
\newcommand{\BH}{black hole}
\newcommand{\BHs}{black holes}
\newcommand{\EM}{electromagnetic}

\newcommand\eg{{\it{e.g.}}}

\newcommand\lo{\mathrel{\raise.3ex\hbox{$<$}\mkern-14mu\lower0.6ex\hbox{$\sim$}}}
\newcommand\go{\mathrel{\raise.3ex\hbox{$>$}\mkern-14mu\lower0.6ex\hbox{$\sim$}}}

\begin{document}
\title{Production of  axions during  scattering of \Alfven waves by fast-moving  \Sch\  black holes
}

\author{Maxim Lyutikov\\
Department of Physics and Astronomy, Purdue University, \\
 525 Northwestern Avenue,
West Lafayette, IN
47907-2036 }

\begin{abstract}
We discuss a novel mechanism of axion production during 
scattering of \Alfven waves by a fast  moving 
 \Sch\  \BH.  The process couples classical macroscopic objects,  and  effectively large   amplitude  \EM\  (EM)  waves,  to  microscopic axions. The key ingredient is that the motion of a \BH\  (BH)  across \Bf\ creates  classical  non-zero  second Poincare invariant, the electromagnetic anomaly  (Lyutikov 2011).  In the case of  magnetized plasma supporting \Alfven wave, it is the fluctuating component of the \Bf\  that contributes to the anomaly: for sufficiency small BH   moving with the super-Alfvenic velocity the  plasma does not have enough time  to screen  the  parallel \Ef. This 
 creates  time-dependent  $ \E \cdot \B  \neq 0$, and production of axions via the  axion-EM coupling.
  \end{abstract}

\maketitle

\section{Introduction}

Axions are   hypothetical  particle invoked  to resolve the strong CP problem in quantum chromodynamics  \cite{2010RvMP...82..557K}. They are also invoked as dark matter candidates \cite{2014ChPhC..38i0001O}.

The present model  of axion production is based  on the observation that a \Sch\ BH moving in vacuum across \Bf\ generates non-zero  second Poincare invariant  $\E \cdot \B \neq 0$, Ref. \citep{2011PhRvD..83f4001L}.  This effect, combined with the idea of  axion-photon mixing \cite{1983PhRvL..51.1415S}, particularly  in the presence  of external  magnetic fields \cite{1988PhRvD..37.1237R},  may lead to axion production  via the    EM anomaly.

 Consider an   ideal plasma with density $n$  in external \Bf\ $\B= B_0 {\bf e}_z$. The plasma supports  \Alfven waves, low frequency oscillations of the \Bf, with  frequency of the waves $\om_A= v_A k$  smaller than the plasma  frequency  $\om_p$, $\om_A \leq \om_p$.
Here 
\be
 v_A = \frac{B_0}{\sqrt{ 4\pi n  m_p}}
 \label{vA}
 \ee
 is  \Alfven velocity, $B_0$ is the \Bf, $m_p$ is proton mass. 
  \Alfven waves create fluctuating transverse components of the \Bf\ $\delta \B$. 

Next, let a BH move along the  \Bf\ with velocity $ \beta_{BH} c $. Let $ \beta_{BH} c \gg v_A$, so that the BH  moves through  nearly stationary  wiggled \Bf. In the frame of the BH the \Alfven wave is seen as a propagating wave with frequency $\om= \beta_{BH} k  c $. For sufficiently high velocity of the BH the frequency $\om$ may be larger than the electron plasma frequency 
\be
\om_p =\sqrt{\frac{4 \pi n e^2}{m_e}} 
\label{omp}
\ee
(the corresponding conditions are  discussed in \S \ref{Mpp}). 
As a result, the plasma does not have time to adjust to the  $\E\cdot\B =0$ condition: the BH sees a nearly vacuum \EM\ wave, but propagating with $ \beta_{BH}\ll 1$. In the wave the fluctuations of the \Ef\ are much smaller than of the \Bf, and can be neglected (in the wave $\delta  E/ \delta B \sim v_A/c\ll 1$  for non-relativistic \Alfven velocity)

Thus, a  BH moves through a spatially varying, nearly vacuum \Bf\ $\delta \B$, directed perpendicular to the direction of BH motion. Motion of the BH through \Bf\ in vacuum generates non-zero  second Poincare invariant  $\E \cdot \B \neq 0$, Ref. \citep{2011PhRvD..83f4001L}. 
In the frame of the BH it is varying with frequency $\om$.  Axions  with mass $m_A \sim  (\hbar/c^2)  \om $ are then produced by coupling to the \EM\ anomaly. 

\section{The model}
\label{Themodel}

\subsection{Static anomaly: motion of BH across \Bf\ in vacuum}

First we highlight the basic ingredient of the model, that motion of a BH through \Bf\ in vacuum generates non-zero  second Poincare invariant  $\E \cdot \B \neq 0$, Ref. \citep{2011PhRvD..83f4001L}. 

Consider \Bf\ along $z$ direction and a \Sch\ BH moving along $x$ direction.
Using standard relations for \EM\ fields in general relativity \cite{MTW,LLII,1989pbh..book.....N} and
choosing the four-potential in Schwarzschild coordinates 
\ba &&
A_0 = \alpha   \beta_{BH} r \sin \theta \cos \phi B_0
\nn &&
A_\phi = \frac{1}{2} r \sin \theta B_0
\nn &&
\alpha =\sqrt{1 - 2 M_{BH} /r}
\label{4A}
\ea
(in unites $c=G=1$; $\theta$ is the polar angle, $\phi$ is the azimuthal angle), 
we find the EM tensor
\ba &&
F^{\mu\nu} =
\left(
\begin{array}{cccc}
 0 & 0 & 0 & 0 \\
 0 & 0 & -\cos (\theta ) \cos (\phi ) \beta _0 & \alpha \sin (\theta )+\sin
   (\phi ) \beta _0 \\
 0 & \cos (\theta ) \cos (\phi ) \beta _0 & 0 & \cos (\theta ) \\
 0 & -\alpha \sin (\theta )-\sin (\phi ) \beta _0 & -\cos (\theta ) & 0 \\
\end{array}
\right) B_0
\nn &&
F^{\mu\nu} _{;\nu}=0
\nn &&
{\rm Det} F= \frac{4 \beta _0^2 B_0^4 M^2 \sin ^2(\theta ) \cos ^2(\theta ) \cos ^2(\phi )}{r^2} = (\E \cdot \B)^2
\ea

Explicitly, 
\ba &&
\E= \left\{\sin (\theta ) \cos (\phi ),  \alpha  \cos (\theta ) \cos (\phi
   ),-\alpha \sin (\phi )\right\}  \beta_{BH} B_0
    \nn &&
   \B = \left\{-\cos (\theta ), \alpha \sin (\theta ) ,0\right\} B_0
   \nn &&
   \E\cdot\B = -\frac{M \sin (2 \theta ) \cos (\phi )}{r}  \beta_{BH} B_0^2 \propto   \beta_{BH} B_0^2 \frac{M }{r}
  \label{EdotB}
   \ea
Note that  the anomaly  $\E\cdot\B \neq 0 $ is highly non-local, $\propto 1/r$.
 We also comment that   $\E\cdot\B \neq 0$ does not appear during scattering of an \EM\  wave by the \BH,  Appendix \ref{EMSch}. It is important that the \Alfven waves have  non-vacuum dispersion.
 
There is  a non-zero divergence of the electromagnetic topological current $J_\nu$
\ba &&
J_\nu = A^\mu  ({^*} F_{\mu\nu})
\nn &&
J_0 = {\bf A} \cdot {\bf B}=0
\nn &&
J_i= {\bf E}  \times {\bf A} + { A_0 \over \alpha}  {\bf B}
\nn &&
 J_{\mu;\mu} = -{7\over 4} \sin 2 \theta \cos \phi B_0 E_0 { M \over r} = {7\over 4} {\bf E} \cdot  {\bf B}
 \label{JJ}
\ea
Thus, the  non-zero   second  Poincare  electromagnetic  invariant leads to the appearance of  sources of topological  axial vector currents. This    can lead to the  local violation of the baryon and lepton numbers through the triangle
anomaly   \cite{1976PhRvL..37....8T,1996PhyU...39..461R}.

 \subsection{Motion of black hole through \Alfven wave creates EM anomaly}
 
 Consider  an \Alfven wave with the wave number $k$ and fluctuating \Bf\ $\delta \B$.
 In the frame of the BH moving non-relativistically with $\beta_{BH}  c \gg v_A$, the transverse component of the \Bf\ varies as 
 \be
 \delta B(t)= \delta B \cos( k( \beta_{BH}  t-  z))
 \ee
In  the expression for the anomaly (\ref{EdotB}) we can then put $B \to   \delta B(t)$. 
 \be
  \E\cdot\B  \propto  \beta_{BH}  (\delta B)^2  \cos^2( k( \beta_{BH}  t-  z)) \frac{M }{r}
  \label{EdotB1}
  \ee
 
 The sign of $\E\cdot\B$ depends on the location, $\propto  \sin (2 \theta ) \cos (\phi )$, Eq (\ref{EdotB}).
 
  \subsection{Coupling to axions }
 
 Axions  interact only minimally with ordinary matter.
 Axions couple to EM fields via the anomaly
 \ba &&
 {\cal{L}} _{a\gamma} = g_{a\gamma}  a  (\E \cdot \B )
 \nn &&
  g_{a\gamma}= \xi  \times  2 \times 10^{-10} GeV^{-1} \frac{ m_a}{1 eV}
  \ea
$ a$ stands for axion field,   $m_a$ is axion mass (expected in the range $10^{-6} - 1 $ eV, Ref \cite{2008LNP...741...51R}), $\xi$ is some parameter.
  
  Using (\ref{EdotB1}) we find
  \be
  {\cal{L}} _{a\gamma} = g_{a\gamma}  a   \beta_{BH}  (\delta B)^2  \cos^2( k( \beta_{BH}  t-  z)) \frac{M }{r}
  \ee
  
  Thus, we have  a {\it classical} configuration with time dependent $\E\cdot \B$ anomaly. In this   case the axion production is possible and can be  large (it is proportional to large classical field, rather than small quantum  fluctuations).

The  resonance condition, $\om_a = m_a$,  requires that the Compton length of axions
  \be
  \lambda_a = \frac{\hbar }{m_a c}
  \label{lambdaa} 
  \ee
  matches temporal variations of the anomaly,
  \be
   k _{res}  = \frac{1}{\beta_{BH} \lambda_a } 
   \label{kres}
    \ee

\section{Astrophysical applicability }
 \label{Mpp}
 
 The process discussed in \S \ref{Themodel} is theoretically possible,  but how realistic is it? Conceptually, the main  problem  is that the model invokes macroscopic effects  (which are qualitatively large in value) to produce microscopic particles.  Since the axion production  is resonant, the corresponding scales must match: this is the main limitation/uncertainty.

  Many astrophysical setting are possible, from stellar mass {\BH}s moving through magnetized interstellar medium (ISM), to primordial \BHs, to the processes in the Early Universe. As a basic example (which will be shown to be  hard to satisfy), let us consider a BH moving trough an ISM. The first requirement is that the BH moves faster than \Alfven waves.
 In a typical Galactic  weakly magnetized plasmas the \Alfven velocity  (\ref{vA}) is sub-relativistic;
 typical values in the interstellar medium are $\sim 10-100$ km/sec \cite{1977ApJ...218..148M}.
 If velocity of the BH is smaller than $v_A$, then
 variations of the EM field on the scale of the BH  will occur on time scale $R_g/v_A$, where $R_g =  2 G M/c^2$ is the \Sch\ radius. On the other hand if velocity of the BH is lager than $v_A$, variations of the EM field on the scale of the BH then will occur on time scale $R_g/ \beta_{BH}$. Assuming large \Alfven Mach number  $M_a=  \beta_{BH}  c/  v_A \gg 1$, and equating $ \beta_{BH} c/R_g = \om_p$ we find a mass of the BH so that  the plasma time $1/ \om_p$ equals light travel time over the horizon.
 \be
  M_{BH} = \frac{ c^3 \sqrt{m_e} }{4 \sqrt{\pi} e G \sqrt{n}} \beta _{BH}  = 3.6 \times 10^{33}   \beta _{BH}  n^{-1/2} \,  {\rm gramm}
 \ee
  Thus in plasma of  density $n=1$  cm$^{-3}$, a $\sim$ Solar mass BH  moving subrelativistically has light travel time of the order of plasma time. Smaller BH will  induce $E_\parallel$ that  will  not be screened by plasma.

  High spacial velocity of stellar-mass BHs may come from merger.
It is expected that   
  BH kick during mergers can be as high as $ \beta _{BH} \sim 10^{-2} \sim {\rm few} \, 10^3$ km/s \citep{2007ApJ...659L...5C}.
 Faster moving BH produces shorter time scale variations;  this eases constraints on the condition that plasma effects do not short out $E_\parallel$. 
Thus,  stellar-mass BHs moving with Mach number  $M_A \geq 1$ can produce variations of EM fields on time-scale shorter that plasma scale, and thus, parallel \Ef, and the EM anomaly.

Next, let us estimate  the Compton length of axions, and compare it to the astrophysical expectations (this is needed for the resonant axion production). For    the expected axion mass of $\sim 10^{-6} $ eV, the Compton length of axions (\ref{lambdaa}) evaluates to 
  \be
  \lambda_a =20 \, \left( \frac{m_a}{10^{-6} {\rm eV}}\, \right)^{-1} \, {\rm cm} 
  \label{lllaa}
  \ee
  The BH with \Sch\ radius that equals $\lambda_a$  would have a mass 
    \be
    M_{BH,a} =\frac{ c \hbar}{2 G m_a} = 6 \times 10^{-5} M_\odot  \left( \frac{m_a}{10^{-6} {\rm eV}}\, {\rm cm}  \right)^{-1},
    \ee
    just somewhat larger than the mass of the Earth.
We arrive at an important point: macroscopic objects (of the order of the mass of the Earth) {\it can} couple to the microscopic axions.

The resonant frequency of the anomaly's oscillation in the frame of the BH,   $\om = \beta_{BH}  c   k _{res} $, with   $k _{res} $  given by (\ref{kres}), should be larger than $\om_p$. This requires
\be
n \leq \frac { m_a^2  m_e c^4}{ 4 \pi e^2 \hbar^2 } = 7 \times 10^8 \left(  \frac{m_a}{ 10^{-6} {\rm  eV}} \right)^2 {\rm cm}^{-3},
\ee
an easily satisfiable condition.

Another  constraint comes from the condition that before the BH comes, the plasma must support  fairly short wavelength  \Alfven oscillation (\ref{lllaa}). This requires (at least) 
that $\lambda _a$ be larger than the Debye length  $r_D$,
\ba && 
\frac{  \lambda_a}{r_D} = \beta_{BH}^{-1}
 \frac{ 2 \sqrt{\pi}  e  \hbar \sqrt{n}}{c m_a \sqrt{ k_B T} } =30 \times  \left(  \frac{m_a}{ 10^{-6} {\rm  eV}} \right) n^{-1/2}  \geq 1
 \nn &&
 n \leq \frac{ (m_a c)^2 k_B T}{ 4\pi e^2 \hbar^2 \beta_{BH}^2} = 10^3   \left(  \frac{m_a}{ 10^{-6} {\rm  eV}} \right)^2  \beta_{BH}^{-2} \,  {\rm cm}^{-3}
\nn &&
r_D= \frac{v_T}{\om_p}
\ea
where $T$ is the temperature of the ISM plasma.
Also a satisfiable condition. 

The  most  stringent constraint comes from the fact that resistive effects in the ISM dissipate  short  wavelength \Alfven waves. 
For a BH moving with $\beta_{BH}$ through an \Alfven  wave of  wavelength $\lambda_A $ the frequency of oscillations seen in its frame,  $ \beta_{BH} c/ \lambda   $, should match the axion mass (Eqns (\ref{lambdaa}-\ref{kres}): 
\be
\lambda_A  = \frac{2 \pi}{k_{res}} = 10^3 \, \beta_{BH}  \left( \frac{m_a}{10^{-6} {\rm eV}}\, {\rm cm}  \right)^{-1} \,  {\rm cm} 
\ee
This wavelength is substantially smaller that the expected  inner  scale of Kolmogorov turbulence 
in the ISM, $ l_{min} \sim 10^{10}$ cm, Ref.  \cite{1995ApJ...443..209A}. 
Thus,  a single stellar mass BHs moving through ISM are not like to encounter \Alfven waves of the required properties (see  below a comment on the beat oscillations of the anomaly between two BHs).

    \section{Discussion}
    
    The proposed mechanism of axion production involves combined effects of  several physics disciplines: electromagnetism and plasma physics (\Alfven waves), General Theory of Relativity (Black holes),   and particle physics (axions). The proposed mechanism couples large classical  quantities to the weakly interacting axions, hence can be highly efficient.
      The mechanism is somewhat  related  to  axion-photon mixing in magnetic fields \cite{1988PhRvD..37.1237R}.

      The proposed    mechanism of axion production has a number of  specific points/advantaged:
    \begin{itemize}
    \item it involves interaction of macroscopic classical fields and matter (hence could be more powerful than typically small quantum effects). 
    \item it involves macroscopic fluctuating \EM\ fields  (whose values are typically much larger than that of the photon fields). As an example, consider an \Alfven wave in \Bf\ $B_0$ with relative amplitude $a_H \equiv \delta B/B_0 \leq 1$. The laser non-linearity parameter  \cite{1975OISNP...1.....A} then estimates to 
    \be
    a \equiv \frac{ e (\delta B) }{m_e c \om}  = a_H \frac{e B_0}{m_e c^2 k \beta_{BH} }=
    2  a_H \frac{e  G M_{BH} B_0}{ m_e c^4  \beta_{BH} }= 10^2 a_H  \beta_{BH}  ^{-1} B_0  \left( \frac{M_{BH}}{M_\odot}  \right) 
     \ee
     where in the last relation we estimated $k \sim 1/R_g$; $B_0$ is in Gauss. This is incredibly intense EM wave, far beyond what is reachable in the laboratory experiments.
     
    \item it is non-local (so that different regions produce axions incoherently, this eliminates strong  cancellation).   One further complication involves interference between newly produced axions. If axions are produced with typical velocity $v_a \ll c$, their de Broglie wavelength (coherence scale)  $\lambda_{D,a} \sim 1/(m_a v_a)$ is much larger than  the Compton length  $\lambda _a$, Eq. (\ref{lambdaa}), Refs \cite{2007PhRvL..98q2002S,2019PhRvD..99b3015L,2021arXiv210708040M}. The coherence scale $\lambda_{D,a}$   depends on environment and variations of the surrounding fields, 

    \end{itemize}
    
    Moving Kerr BH adds another level of complexity \cite{2014PhRvD..89j4030M}. First, Wald's solution \cite{Wald} also produces (stationary) $\E \cdot \B \neq 0$. In an \Alfven 
    wave (\eg,  polarized orthogonally to the BH's spin)  the anomaly will be time-dependent  and thus can couple to axions. Fast motion  of a BH along the \Bf\ then can ensure that the corresponding variations are sufficiently fast and not screened by plasma. 
    
Our estimates demonstrate that a single (sub)stellar mass BHs moving in an  ISM is not likely to produce axions, due to the lack of resonant \Alfven waves (\Alfven waves of the required wavelength   of $\sim$ a meter can  marginally  propagate  in the ISM plasma, but suffer quick, on astronomical time scales, resistive decay). 

 If there are many BHs moving uncorrelatedly, then, on the one hand,  the $\E \cdot \B$ will tend to average out on  scales much larger than the typical separation between BHs  (but locally it will still be dominated by a single one, since $\E \cdot \B \propto  1/r$). On the other hand, two BHs with somewhat different velocities will produce  the anomaly varying on the beat frequencies. The minus-beat frequency can be much smaller, and couple to longer wavelengths \Alfven waves. 
We leave investigation of other possible astrophysical sites  (\eg\ \Alfven waves propagating in \mss\ of compact objected with strong gravity)  to a future work.

      \section{Acknowledgements}

This work is  supported by   NSF grants 1903332 and  1908590. 
I would like to thank  Robert Brandenberger, Sergey Khlebnikov, Martin Kruczenski and particularly Ariel Zhitnitsky for discussions.

   \bibliographystyle{apsrev}
\bibliography{/Users/maxim/Home/Research/BibTex}

\appendix

\section{EM waves in Schwarzschild spacetime}
\label{EMSch}

Using spherical wave functions, we can express the EM four-vector \cite{MTW}
\be
A_\mu = \left\{ 0, \frac{f(r)}{r^2} Y_{lm}, \frac{g(r )}{ l(l+1) }\frac{ \partial_r f}{r} \partial_\theta  Y_{lm},  \frac{g(r)}{ l(l+1)} \frac{ \partial_r f}{r \sin \theta} \partial_\phi  Y_{lm}\right\} 
\ee

We find equations for $f$ and $g$,
\ba &&
2 r \alpha (\alpha - g) f'-(1-\alpha^2) f=0
\nn &&
 r^3 \alpha^3\partial_r ( g f')  -\left( l(l+1)  \alpha^2 - r^2 \om^2\right) f =0
 \label{fg}
 \ea
 Or
 \ba &&
 g= \alpha - \frac{( 1-\alpha^2) f}{2 r \alpha f}
 \nn &&
 f^{\prime \prime} =  \left( \frac{ l(l+1) }{\alpha^2 r^2} - \frac{-3 \alpha^4+2 \alpha^2+4 r^2 \omega ^2+1}{4 \alpha^4 r^2} \right) f
\ea
 In flat space Eq. (\ref{fg}) give  $g(1)=1$, $f= j_{l}( \om r) = \sqrt{r} J_{l+1/2} (\om r)$, where $ j_{l}$ are spherical Bessel functions.  

Importantly, ${\rm Det} F_{\mu \nu} =0$ in this case:  there is no EM  anomaly.

\end{document}